\documentstyle[pra,aps]{revtex}

\twocolumn

\begin{document}

\title{Nonlinear energy-loss straggling of 
protons and antiprotons in an electron gas}
\author{Neng-ping Wang$^{\dagger}$}
\address{Departamento de F\'\i sica de Materiales, Facultad de Ciencias Qu\'\i
micas,\\ Universidad del Pa\'\i s Vasco/Euskal Herriko Unibertsitatea,\\ 
     Apartado 1072, San Sebasti\'{a}n 20080, Spain \\}
\author{J. M. Pitarke}
\address{Materia Kondentsatuaren Fisika Saila, Zientzi
Fakultatea, Euskal Herriko  Unibertsitatea,\\ 644 Posta kutxatila, 48080 Bilbo,
Basque Country, Spain}
\date{12 November 1997}
\maketitle

\begin{abstract}
The electronic energy-loss straggling of protons and antiprotons moving at
arbitrary nonrelativistic velocities in a homogeneous electron gas are evaluated
within a quadratic response theory and the random-phase approximation (RPA).
These results show that at low and intermediate velocities quadratic
corrections reduce significantly the energy-loss straggling of
antiprotons, these corrections being, at low-velocities, more important than in
the evaluation of the stopping power.
\end{abstract}


\narrowtext

The energy loss of charged particles has received a great deal of attention
for several decades\cite{Bohr,Bethe,Bloch,Fermi,Lindhard}, since it plays an
important role in investigating the elemental composition, depth distribution,
and lattice location of implanted atoms in matter. The characterization of the
distribution of energy losses suffered by ions in their interaction with matter
requires, in the simplest case, two quantities: the stopping power and the 
energy-loss straggling. 
Calculations of the energy-loss straggling of charged particles in an electron
gas were performed\cite{Bonderup}, within linear response
theory, by treating the screened interaction to lowest order in the projectile
charge ($Z_1e$). This procedure leads results for the stopping power and the
energy-loss straggling that are proportional to $Z_1^2e^2$. However, early
measurements by Barkas et al.\cite{Barkas} on positive and negative pions, and
also subsequent experiments with protons and antiprotons\cite{Andersen}, have
indicated that these quantities exhibit a dependence on the sign of the
projectile charge. This dependence was ascribed to
$Z_{1}^{3}$ corrections to the
first Born approximation that underlies the linear response theory, assuming
that higher odd power corrections were negligible
in the velocity regime under study. Very recent measurements\cite{Moller}
of the  stopping power for antiprotons in light and heavy targets have shown
that the stopping power for antiprotons is
reduced, near the stopping maximum, significantly (about 35\%) as compared to 
the corresponding stopping power for protons.

The $Z_1^3$ correction to the stopping power was first calculated by using
classical perturbation theory for a harmonic oscillator\cite{Ashley0}, and
quantal calculations of the $Z_1^3$ correction to the stopping power of an
electron gas have also been performed, in the low\cite{Hu} and
high\cite{Sung,Esbensen} velocity limits. More recently, a
quadratic response theory for the interaction of charged particles with an
electron gas has been developed\cite{Pitarke2}, in which full account of the
Fermi motion of the target is taken into account, and calculations, within this
theory, of the stopping power of an electron gas for ions moving with arbitrary
non-relativistic velocities\cite{Pitarke1} have provided good agreement with
measurements of the energy loss of protons and antiprotons in
silicon\cite{Andersen}. A calculation of the $Z_1^3$ correction to the
energy-loss straggling and the energy width of the states of ions moving in an
electron gas has also been reported, in the low-velocity limit, and it has been
demonstrated\cite{Pitarke2,Wang} that this correction to the quantities that
characterize the distribution of electronic energy losses, i.e., the stopping
power, the energy-loss straggling and the energy width coincides, in this limit,
with the result of full nonlinear density-functional
calculations\cite{Echenique1,Ashley,Nagy}, for high electron densities and
small projectile charges.

In this paper we investigate, within the full random-phase approximation (RPA),
the $Z_1^3$ correction to the energy-loss straggling of protons and antiprotons
moving with arbitrary nonrelativistic velocities in an electron gas. The electron
gas model is well-known to account for the contribution to both the stopping power
and the energy-loss straggling coming from the interaction of the projectile with
valence electrons in metals\cite{Echenique2}, and the total contribution to these
quantities coming from all target electrons has also been computed, on the basis of
the electron gas model of the target, by using the so-called local plasma
approximation\cite{Lindhard}. The results throughout this paper will  be expressed in
Hartree atomic units, i.e.,
$\hbar=m=e=1$. 

We consider an ion of charge $Z_{1}$ moving with constant velocity ${\bf v}$ 
through an isotropic homogeneous electron gas embedded in a uniformly
distributed positive background. The energy-loss straggling per unit path
length of the projectile, ${\rm d}\Omega^{2}/{\rm d}x$, can be expressed in
terms of the probability  of transferring momentum
$\bf q$ and energy $\omega$ to the electron gas, $P({\bf q},\omega)$,
as\cite{Echenique2}
\begin{equation}
{{\rm d}\Omega^2\over{\rm d}x}={1\over v}\int d{\bf
q}\int_{0}^{\infty}\omega^{2} P({\bf q},\omega){\rm d}\omega.
\end{equation}
An explicit expression for this probability is given, up to third
order in the  projectile charge, in Ref.\onlinecite{Pitarke2}, by following a 
perturbation-theoretical analysis of the many-body interactions between the 
moving charge and the electron gas. Using this probability one finds, within
the RPA, from Eq.(1): 
\begin{equation}
{{\rm d}\Omega^2\over{\rm d}x}=\left({{\rm
d}\Omega^2\over{\rm d}x}\right)_L+\left({{\rm d}\Omega^2\over{\rm d}x}\right)_Q,
\end{equation}
where $\left({{\rm
d}\Omega^2/{\rm d}x}\right)_L$ and $\left({{\rm
d}\Omega^2/{\rm d}x}\right)_Q$ represent contributions to the energy-loss
straggling that are proprotional to $Z_1^2$ and $Z_1^3$, respectively. $\left({{\rm
d}\Omega^2/{\rm d}x}\right)_L$ represents the well-known result one obtains within
linear response theory\cite{Echenique2}:
\begin{equation}
\left({{\rm d}\Omega^2\over{\rm d}x}\right)_L=
2{Z_1^2\over v}\int{{\rm d}^3{\bf q}\over (2\pi)^3}\omega^{2}
            {\cal V}_{\bf q}{\rm Im}(-\epsilon_q^{-1})
\theta(\omega),
\end{equation}
and $\left({{\rm
d}\Omega^2/{\rm d}x}\right)_Q$ appears as a
consequence of the so-called quadratic response of the electron gas:
\begin{eqnarray}
\left({{\rm d}\Omega^2\over{\rm d}x}\right)_Q&=&-4{Z_1^3\over v}\int{{\rm
d}^3{\bf q}\over (2\pi)^3}\omega^{2}{\cal V}_{\bf q}\theta(\omega)
\int{{\rm d}^3{\bf
q}_1\over (2\pi)^3}{\cal V}_{{\bf q}_{1}}\cr
&\times&{\cal V}_{{\bf q}-{\bf q}_{1}}
\left[f_1(q,q_1)+f_2(q,q_1)+f_3(q,q_1)\right],
\end{eqnarray}
with
\begin{equation}
f_{1}(q,q_1)={\rm Im}\epsilon_q^{-1}{\rm Re}\epsilon_{q_1}^{-1}{\rm
Re}\epsilon_{q-q_1}^{-1}{\rm Re}M^{s}_{q,q_1},
\end{equation}
\begin{equation}
f_{2}(q,q_1)={\rm Re}\epsilon_q^{-1}{\rm Re}\epsilon_{q_1}^{-1}{\rm
Re}\epsilon_{q-q_1}^{-1}H^{s}_{q,q_1},
\end{equation}
and
\begin{equation}
f_{3}(q,q_1)=-2{\rm Im}\epsilon_q^{-1}{\rm Im}\epsilon_{q_1}^{-1}{\rm
Re}\epsilon_{q-q_1}^{-1}H^{s}_{q,q_1}.
\end{equation}
Here, $q=({\bf q},\omega)$, $q_1=({\bf q}_1,\omega_1)$, ${\cal V}_{\bf q}$ is
the Fourier transformation of the electron-electron bare Coulomb interaction,
\begin{equation}
{\cal V}_{\bf q}={4\pi\over{\bf q}^2},
\end{equation}
$\epsilon_q$ is the longitudinal dielectric function of the medium,
$M_{q,q_1}^s$ represents the quadratic response function, $H_{q,q_1}^s$ is
related to the imaginary part of $M_{q,q_1}^s$, as shown in
Ref.\onlinecite{Pitarke2}, $\theta(x)$ is the Heaviside function, $\omega={\bf
q}\cdot{\bf v}$, and $\omega_1={\bf q}_1\cdot{\bf v}$.

We have calculated linear and quadratic contributions to the electronic
energy-loss straggling for a wide range of non-relativistic velocities of the
projectile, after substitution of the full RPA linear and quadratic response
functions into Eqs. (3) and (4). The result of this calculation for an electron
density parameter $r_s=2$ ($r_s=[3/4\pi n]^{1/3}$, $n$ being the electron
density) and $Z_1=1$ is presented in Fig. 1, as a function of the velocity of 
the projectile, by dashed (linear contribution) 
and solid (quadratic contribution)
lines, together with the total (linear + quadratic) energy-loss straggling of
protons (dashed-dotted line) and antiprotons (dotted line).
It is obvious from this figure that at low and intermediate velocities quadratic
contributions result in a significant reduction of the energy-loss straggling of
antiprotons. On the other hand, though at low velocities the $Z_1^3$ effect
appears to be more important in the energy-loss straggling than in the
evaluation of both the stopping power and the energy-width, at high velocities
the quadratic contribution to the energy-loss straggling decreases very rapidly.

The numerical results of the total RPA $Z_1^3$ contribution to the energy-loss
straggling for different values of the electron density parameter $r_s$ are
illustrated in Fig. 2. At low velocities, the $Z_1^3$ term gets larger as the
electron density increases, as discussed in Ref.\onlinecite{Wang}. At high
velocities, $Z_1^3$ corrections to both the stopping power and the energy-loss
straggling present similar dependences on the electron density.

For completeness, the low-velocity limit of the energy-loss straggling of 
protons and antiprotons is presented in Fig. 3, as a function of the electron
density parameter $r_s$, as obtained up to third order in the projectile charge
after substitution of the low-frequency forms of both linear and quadratic
response functions into Eqs. (3) and (4). Since exchange and correlation are
absent in our RPA treatment, we have also represented, in the same figure, full
nonlinear self-consistent Hartree calculations like the ones presented in
Ref.\onlinecite{Wang}. This figure indicates that in the case of antiprotons 
with $r_{s}\stackrel{<}{\sim}2$ our quadratic 
response calculations agree, in the
low velocity limit, with the corresponding full nonlinear results;
accordingly, our calculations are expected to give, for these
electron densities, good account of the full nonlinear energy-loss straggling of
antiprotons moving at arbitrary velocities. In the case of protons our quadratic
response calculations appear to be accurate, in the low velocity limit, only at
high electron densities ($r_{s}\stackrel{<}{\sim}1.5$).

The results depicted in Figs. 1, 2 and 3 indicate that the $Z_1^3$
contribution to the energy-loss straggling in an electron gas is positive for
$Z_1>0$, for all velocities of the projectile and all electron densities of the
target. As a result, the energy-loss straggling is, within quadratic response
theory, greater for a proton than for an antiproton, as in the case of the stopping
power. This result is, at high electron densities, in agreement with full
nonlinear self-consistent Hartree calculations, as discussed in
Ref.\onlinecite{Wang}. At low electron densities self-consistent Hartree
calculations still predict, in the low-velocity limit, nonlinear effects to
reduce the energy-loss straggling of antiprotons. However, the formation around
protons of bound states, not described within perturbation theory, tends to
screen out interactions with the electron gas, and this causes, at low electron
densities, a significant reduction in the energy-loss straggling of protons, as
can be inferred from Fig. 3. This reduction can be accounted for through the
introduction of an effective charge, as discussed in
Refs.\onlinecite{Echenique1} and\onlinecite{Ashley}.    

In summary, we have calculated $Z_{1}^{3}$ corrections to the energy-loss straggling 
of protons and antiprotons moving at arbitrary nonrelativistic velocities 
in a homogeneous electron gas, by using a quadratic response theory and
the RPA. At high velocities (after the plasmon threshold), the
quadratic response theory provides an accurate estimate of the full nonlinear 
correction to the energy-loss straggling of protons and anitprotons 
moving in an electron gas. In the case of antiprotons moving at arbitrary
nonrelativistic velocities in many solids used in the experiments
($r_{s}\stackrel{<}{\sim}2$), the present quadratic response theory provides an
accurate description  of nonlinearities in the energy-loss straggling. 
A comparison with results obtained within linear response theory indicates
that at low and intermediate velocities the $Z_{1}^{3}$ correction reduces 
significantly the energy-loss straggling of antiprotons, this correction being,
at low velocities, more important than in the evaluation of the stopping power.

\acknowledgements

The authors gratefully acknowledge P. M. Echenique and A. Arnau for useful
discussions. The work of one of us (N. P. W.) was supported by
Departamento de Educaci\'{o}n, Universidades e Investigaci\'{o}n del Gobierno
Vasco (Ref.: 1/97023). We wish also to acknowledge the support of the 
University of the Basque Country and the Basque Unibertsitate eta 
Ikerketa Saila under contracts UPV063.310-EA056/96 and GV063.310-0017/95,
respectively.

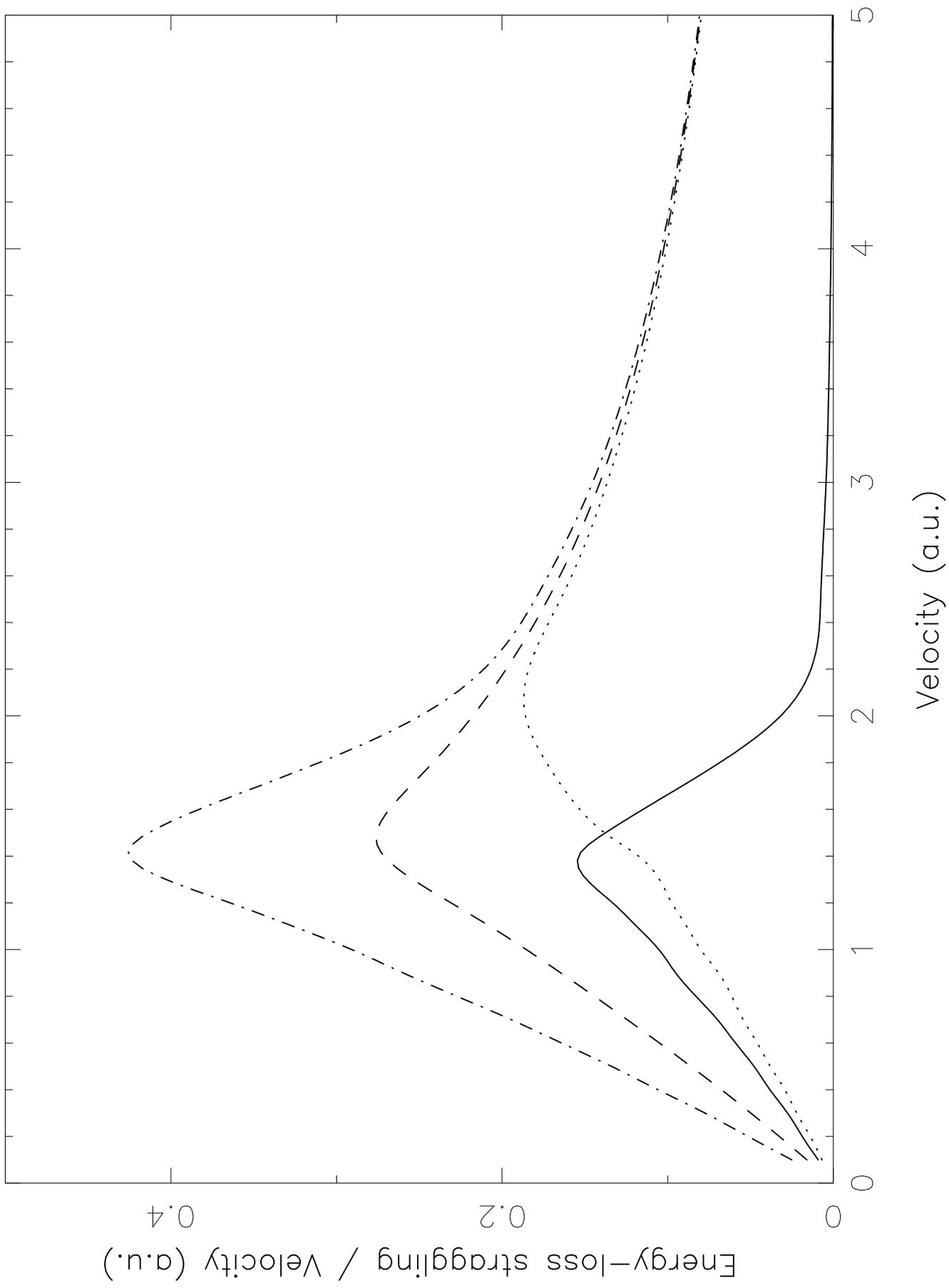
\begin{figure}
\caption{Linear (dashed line) and quadratic (solid line) contributions to the
energy-loss straggling per unit path length for $Z_1=1$ and $r_s=2$, 
divided by the velocity of the projectile and as a function of the velocity. 
Total (linear + quadratic) energy-loss stragglings of protons and antiprotons,
divided by the velocity, are represented by dashed-dotted and dotted lines, 
respectively.}
\end{figure}

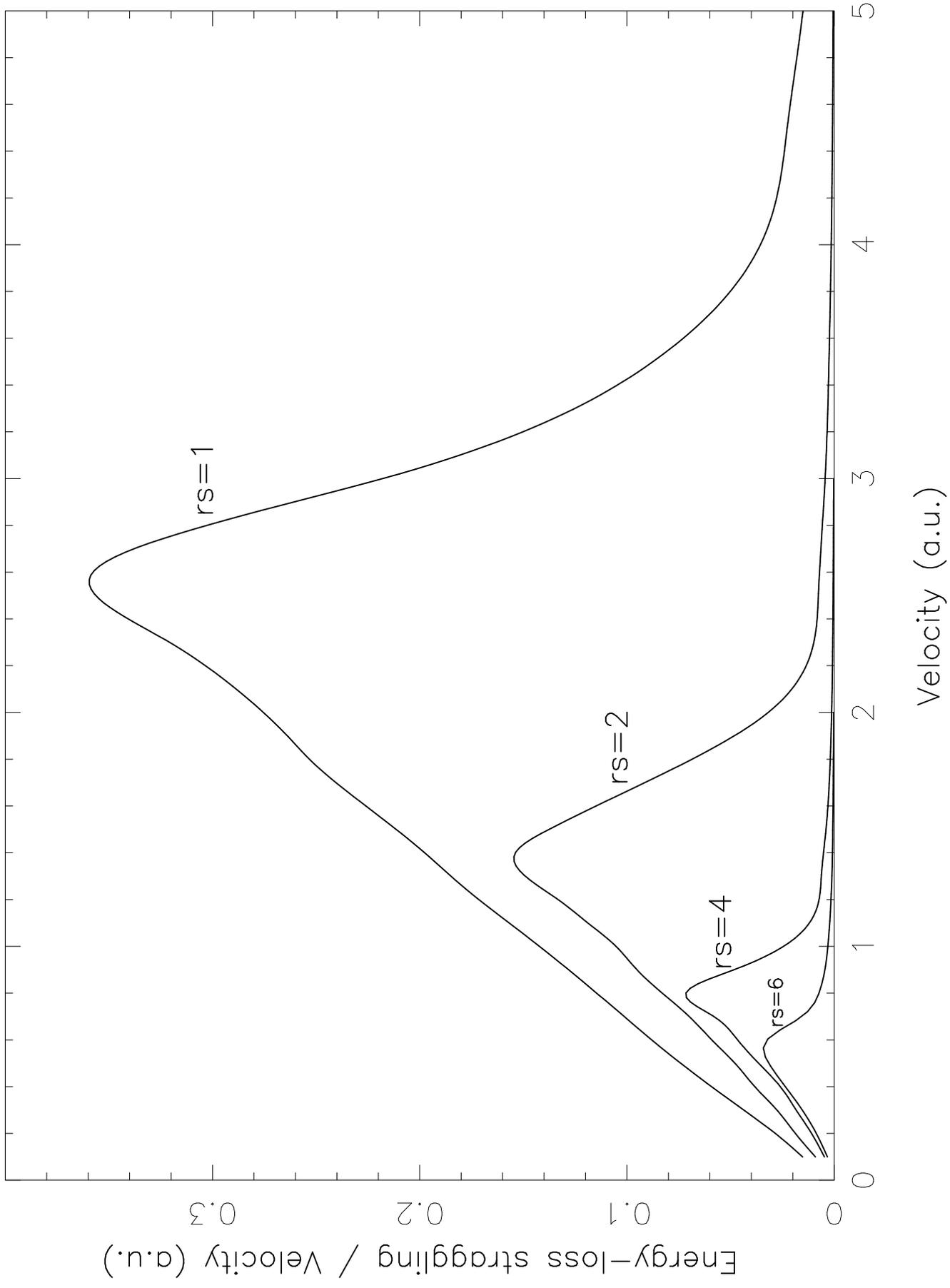
\begin{figure}
\caption{Quadratic ($Z_1^3$) contribution to the energy-loss straggling
per unit path length, divided by the velocity of the projectile and as 
a function of the velocity,
for $Z_1=1$ and four representative values of $r_s$: $r_s=1$, $r_s=2$, $r_s=4$,
and $r_s=6$.}
\end{figure}

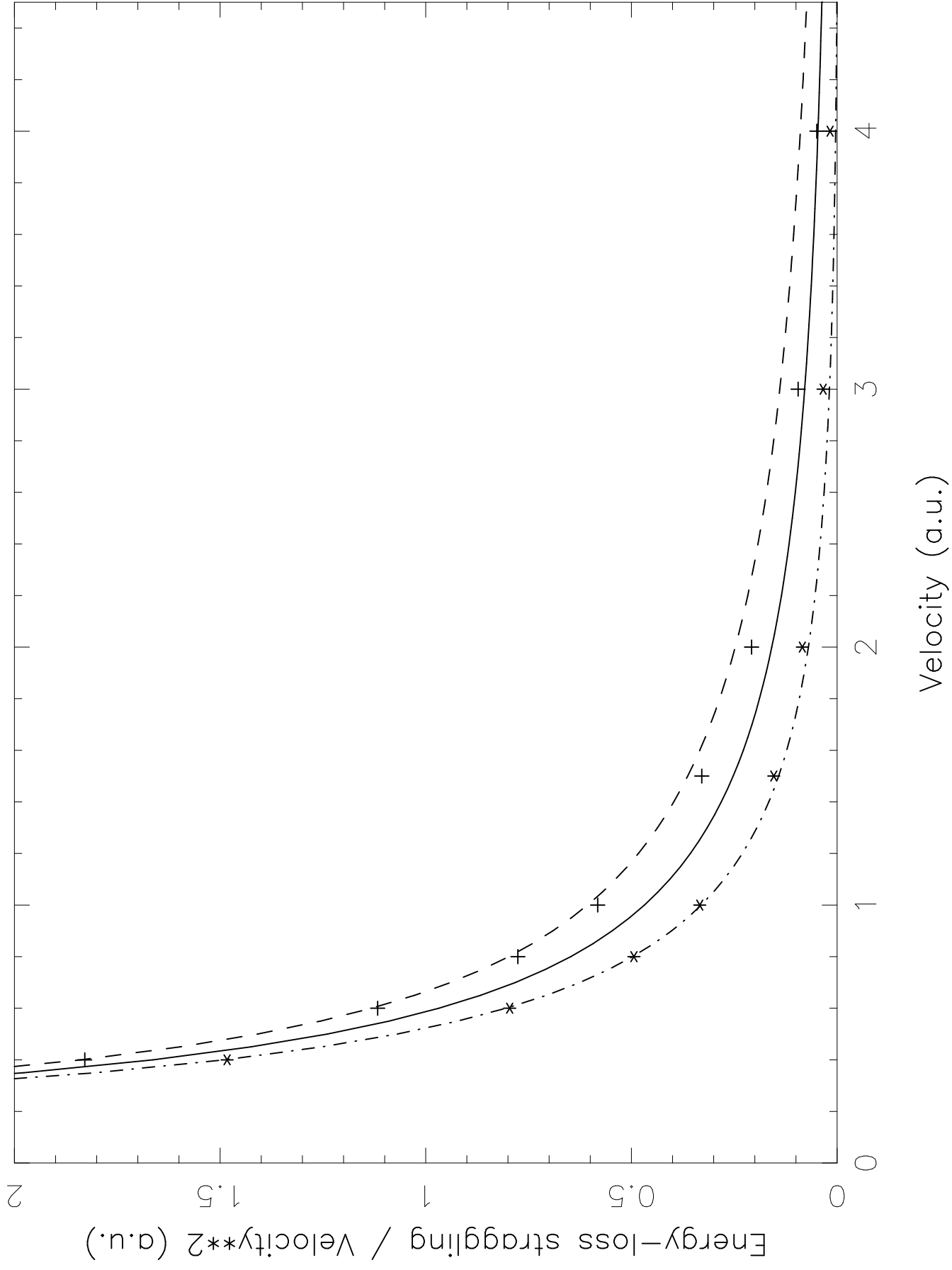
\begin{figure}
\caption{Total (linear + quadratic) energy-loss straggling 
per unit path length divided by the square
of the projectile velocity, as a function of the electron density parameter
$r_s$, for protons (dashed line) and antiprotons (dashed-dotted line).
Crosses and stars represent the full nonlinear energy-loss straggling of
protons and antiprotons, respectively. The solid curve represents the linear
contribution to the energy-loss straggling for $Z_1=1$.}
\end{figure}

\end{document}